\documentclass[acmsmall,screen,nonacm]{acmart}

\usepackage{layout}
\usepackage{varwidth}
\usepackage{tikz}
\usepackage{cleveref}
\usepackage{subcaption}
\usetikzlibrary{quantikz2,shapes.geometric,arrows,positioning}
\DeclareFontEncoding{LS2}{}{}
\DeclareFontSubstitution{LS2}{stix2}{m}{n}

\DeclareMathOperator{\trace}{trace}
\DeclareSymbolFont{integrals}{LS2}{stix2cal}{m}{n}
\DeclareMathSymbol{\defeq}{\mathrel}{integrals}{"85}

\newcommand{\optlev}{\texttt{optimization\_level}}
\graphicspath{{./figures/}}

\newcommand{\Fid}{\mathcal{F}}

\usepackage[]{hyperref}

\begin{document}

\title{COGNAC}
\subtitle{Circuit Optimization via Gradients and Noise-Aware Compilation}

\author{Finn Voichick}
\orcid{0000-0002-1913-4178}
\affiliation{
  \institution{University of Maryland}
  \city{College Park}
  \country{United States}
}
\email{finn@umd.edu}

\author{Leonidas Lampropoulos}
\orcid{0000-0003-0269-9815}
\affiliation{
  \institution{University of Maryland}
  \city{College Park}
  \country{United States}
}
\email{leonidas@umd.edu}

\author{Robert Rand}
\orcid{0000-0001-6842-5505}
\affiliation{
  \institution{University of Chicago}
  \country{United States}}
\email{rand@uchicago.edu}

\begin{abstract}
We present COGNAC, a novel strategy for compiling quantum circuits based on numerical optimization algorithms from scientific computing. Observing that shorter-duration ``partially entangling'' gates tend to be less noisy than the typical ``maximally entangling'' gates, we use a simple and versatile noise model to construct a differentiable cost function. Standard gradient-based optimization algorithms running on a GPU can then quickly converge to a local optimum that closely approximates the target unitary. By reducing rotation angles to zero, COGNAC removes gates from a circuit, producing smaller quantum circuits. We have implemented this technique as a general-purpose Qiskit compiler plugin and compared performance with state-of-the-art optimizers on a variety of standard benchmarks. Testing our compiled circuits on superconducting quantum hardware, we find that COGNAC's optimizations produce circuits that are substantially less noisy than those produced by existing optimizers. These runtime performance gains come without a major compile-time cost, as COGNAC's parallelism allows it to retain a competitive optimization speed.
\end{abstract}

\maketitle 

\section{Introduction}

Compiling a quantum program can involve a number of intermediate representations but usually results in the construction of a quantum circuit: a sequence of quantum logic gates at an abstraction level comparable to assembly languages in classical computing~\cite{voqc,qasm}.
Quantum circuit optimizers, like their classical counterparts, employ rewrite rules to shrink these circuits, and recent research in this area has seen impressive developments, both in the number of such equivalences and the strategies to apply them~\cite{queso,quarl}.
While this basic \textit{rewriting} scheme is sensible and familiar to computer scientists, it differs significantly from the \emph{numerical} optimizations that physicists perform on the same quantum systems.

Quantum hardware specialists work at a lower level of abstraction, dealing directly with hardware channels in the language of microwave pulses.
Quantum optimal control theory has progressed significantly in recent years, manipulating pulse schedules and implementing the gates that quantum programmers take for granted~\cite{qoc}.
Optimization at this level must acknowledge two key features of quantum systems that logic gates tend to abstract away: control is both \emph{noisy} and \emph{continuous}.

Noise is a defining characteristic of the noisy intermediate-scale quantum era (NISQ), which describes the current state of quantum computing~\cite{nisq}.
Precise modeling of atomic interactions is notoriously difficult, and unintended side effects are not negligible.
Optimal pulse engineering is a matter of \emph{minimizing} (not \emph{eliminating}) this noise, and the computational tools employed to solve this optimization problem differ significantly from those employed in traditional compilers. 
The control parameters at this level are real-valued and continuous, meaning that any pulse-level instruction can be divided and subdivided (for example, by halving the amplitude in a microwave pulse schedule).
Pulse engineering has become time-consuming, specialized, and iterative, involving physical modeling, calculus, and experimentation. It is no wonder then that optimizers shy away from this domain, treating the quantum logic gate as a standard abstraction barrier between analog pulse optimization and digital compiler optimization, with physicists on the one side and computer scientists on the other.

\begin{figure}
\begin{subfigure}[t]{0.49\columnwidth}
\begin{align*}
	R_z(\theta) &\defeq \begin{bmatrix} e^{-i\theta/2} & 0 \\ 0 & e^{i\theta/2} \end{bmatrix}\\
	R_x(\theta) &\defeq \begin{bmatrix} \cos(\theta/2) & -i\sin(\theta/2) \\ -i\sin(\theta/2) & \cos(\theta/2) \end{bmatrix}
\end{align*}
\end{subfigure}
\begin{subfigure}[t]{0.49\columnwidth}
\begin{align*}
	R_{zz}(\theta) &\defeq \begin{bmatrix} e^{-i\theta/2} & 0 & 0 & 0 \\ 0 & e^{i\theta/2} & 0 & 0 \\ 0 & 0 & e^{i\theta/2} & 0 \\ 0 & 0 & 0 & e^{-i\theta/2} \end{bmatrix}
\end{align*}
\end{subfigure}
\Description{Mathematical definitions for the quantum gates $R_z$, $R_x$, and $R_{zz}$.}
\caption{A parameterized gate set}
\label{fig:gateset}
\end{figure}

Recent work has just begun to blur this boundary between analog and digital compilation.
Compilers have been making better use of real-parameterized gates like those in Figure~\ref{fig:gateset}, which generalize the standard gates $X$, $Z$ and $CZ$.
These gates (and the optimizations they enable) are partly responsible for current records in quantum volume performance benchmarks~\cite{qv13}.
Quantum devices can implement these gates by continuously adjusting the real parameters of a pulse envelope~\cite{pranav}.
General-purpose quantum circuit compilers like Qiskit~\cite{qiskit} and TKET~\cite{tket} now allow users to specify the estimated fidelity (accuracy) of the hardware's native two-qubit gates, which is used for approximations when synthesizing general two-qubit gates.
These compilers can then produce a smaller compiled circuit that is not logically equivalent to the original but whose output distribution suffers from less noise, and thus end up implementing the idealized original circuit more faithfully than the original (noisy) circuit itself.

More recently, BQSKit~\cite{bqskit} has drawn further inspiration from continuous numerical optimization techniques.
This compiler partitions a circuit into three-qubit subcircuits, then, for each window, iteratively attempts to remove gates.
It uses the quasi-Newton L-BFGS algorithm~\cite{lbfgs} to adjust continuous parameters on the remaining gates to compensate for the missing one, succeeding if the resulting unitary is within a given tolerance of the original semantics.

We propose further blurring the line between pulse-level (analog) and gate-level (digital) quantum computing, improving compilers with continuous control, noise awareness, \emph{and} iterative calculus-informed optimization techniques.
Accurate modeling of quantum noise is notoriously challenging~\cite{difficult-noise-modeling}, and so we make a major simplification of the noise model of the quantum system.
Rather than precisely modeling the complex interactions that occur within quantum hardware and trying to characterize different types of noise (leakage into higher energy states, T1/T2 decoherence, cross-talk, etc.), we rely on a simplifying assumption: \emph{shorter pulses produce smaller errors}.
Although not as accurate as the noise models typically employed in quantum optimal control, experiments have found it to be a valid approximation in practice~\cite{ionq-fidelity}.
More importantly, it leads to an efficiently differentiable cost function that encourages zero-duration pulses, which can then be eliminated from the circuit.
We thus find a reasonable middle ground between large-scale circuit optimization and experimentally driven pulse engineering, decreasing gate count in an iterative, gradient-driven way at a moderate scale without the need for additional hardware calibrations.

To bridge the gap between pulse optimization and compilers, we developed COGNAC: Circuit Optimization via Gradients and Noise-Aware Compilation.
In the following sections, we detail our strategy and its implementation as a Qiskit compiler plugin, and we then evaluate and compare its performance relative to existing optimizers.
We hope that our technique serves as a useful tool for improving performance on near-term quantum hardware.

\paragraph{Contributions}
This paper presents COGNAC, a general-purpose noise-aware quantum circuit optimizer adapted from lower-level pulse optimization strategies. 
Using gradient-based numerical optimization, COGNAC discovers \emph{approximate} substitutions that are not feasible with traditional rewrite rules, and its strategy is highly parallelizable.
\Cref{sec:technique} gives a high-level overview of our strategy, and \Cref{sec:impl} discusses our implementation.
Evaluated on an NVIDIA GeForce RTX 2080 GPU across a wide variety of benchmark quantum circuits, our implementation of COGNAC typically produces shorter-depth circuits than state-of-the-art optimizers while remaining competitive in optimization time.
Experiments on quantum hardware (\Cref{sec:evaluation}) confirm that COGNAC's \emph{approximately} equivalent output circuits implement the target circuit semantics more faithfully (once we account for hardware errors) than the \emph{exactly} equivalent circuits output by existing optimizers.

\section{Background}

\textit{Quantum computing} is an interdisciplinary area of research that seeks to apply the principles of quantum mechanics---superposition, entanglement, and interference---in a computational setting, in some cases achieving exponential speedup (asymptotically) over the best known classical algorithms.
A quantum bit (\textit{qubit}) replaces the bit as the fundamental unit of quantum information and is mathematically represented as a two-dimensional complex vector space (rather than a two-element set).
The state space of an $n$-qubit system is then a $2^n$-dimensional vector space, with each dimension corresponding to a different string of $n$ bits.
There are numerous ways to physically realize a qubit, such as the polarization of a photon or the spin of an electron.
Although it is an active area of research, the implementations that are currently the most successful tend to encode a qubit through the energy level of either a superconductor or an isolated atom \cite{fred-quantum-architecture-book}.

For a qubit to be useful for quantum computing, it must be possible to manipulate and control it.
Specific details vary with architecture, but the lowest level of hardware control is often a microwave pulse applied to an electrical component.
These pulses correspond to mathematical \textit{rotations} of the quantum state about some axis in the relevant vector space. Typically, these rotations are specified to form a set of discrete \emph{quantum gates}, as in the following set from the standard textbook~\cite{Nielsen2010}:

\begin{figure}[h]
\[
\begin{array}{ccc}
    X \defeq \begin{bmatrix} 0 & 1 \\ 1 & 0 \end{bmatrix} \qquad &
    Y \defeq \begin{bmatrix} 0 & -i \\ i & 0 \end{bmatrix} \qquad &
    Z \defeq \begin{bmatrix} 1 & 0 \\ 0 & -1 \end{bmatrix}  \\[20pt]
    H \defeq \frac{1}{\sqrt{2}}\begin{bmatrix} 1 & 1 \\ 1 & -1 \end{bmatrix} \qquad &
    S \defeq \begin{bmatrix} 1 & 0 \\ 0 & i \end{bmatrix} \qquad &
    T \defeq \begin{bmatrix} 1 & 0 \\ 0 & e^{i\pi/4} \end{bmatrix} 
\end{array}
\]
\begin{gather*}
    \mathit{CNOT} \defeq\begin{bmatrix}
        1 & 0 & 0 & 0 \\
        0 & 1 & 0 & 0 \\
        0 & 0 & 0 & 1 \\
        0 & 0 & 1 & 0
    \end{bmatrix} 
    \qquad
    \mathit{CZ} \defeq \begin{bmatrix}
        1 & 0 & 0 & 0 \\
        0 & 1 & 0 & 0 \\
        0 & 0 & 1 & 0 \\
        0 & 0 & 0 & -1
    \end{bmatrix} 
\end{gather*}
\caption{Gates from the standard Clifford+T gate set}
\label{fig:clifford}
\end{figure}
Here $\mathit{CNOT}$ and $\mathit{CZ}$ are the two-qubit \emph{entangling gates}, one of which is often the only such gate provided by a quantum software package or supported on a quantum computer.

Given that qubit rotations are implemented via continuous-variable microwave pulses, for software or hardware vendors to support a discrete gate set is counterintuitive. Hardware providers have largely come around to this view, with IBM providing the gates $R_z$, $R_x$, and $R_{zz}$ in \Cref{fig:gateset} on their latest machines. Note that $R_z$ implements the $Z,S$ and $T$ gates up to a scalar when given the arguments $\pi, \pi/2$ and $\pi/4$, and $R_{zz}$ can be used similarly to implement a $\mathit{CZ}$ gate. 

Mathematically, ``applying'' these operations to a state vector is a matter of matrix--vector multiplication, and quantum hardware vendors are responsible for precisely calibrating the microwave pulses that correspond to these matrices.

These low-level rotations, or \textit{quantum logic gates}, can be sequenced to form a quantum circuit.
The semantics of such a circuit can be computed by multiplying the matrices of all of the individual gates, though this is impractical for very large circuits, as the dimensionality of the state space grows exponentially with the number of qubits.

Even on a quantum device, executing these programs is costly, and due to high error rates, computation breaks down as more gates are applied. Various optimizers~\cite{voqc,quartz,queso} have been introduced to alleviate this problem but tend to assume a discrete gate set and, therefore, do not take advantage of the opportunities for continuous-valued optimizations.

\section{Technique}
\label{sec:technique}

\tikzstyle{startstop} = [rounded rectangle, draw=black, minimum width=2cm, minimum height=5mm,text centered]
\tikzstyle{process} = [rectangle, draw=black, minimum width=2cm, minimum height=5mm,text centered]
\tikzstyle{arrow} = [thick,->,>=stealth]

\begin{figure}
\begin{tikzpicture}[node distance=24mm]
\node (start) [startstop] {Input};
\node (parameterize) [process, right of=start] {Parameterized};
\node (partition) [process, right of=parameterize] {Partition};
\node (optimize) [process, right of=partition] {Optimize};
\node (prune) [process, below=5mm of optimize] {Prune};
\node (end) [startstop, right of=prune] {Output};
\draw [arrow] (start) -- (parameterize);
\draw [arrow] (parameterize) -- (partition);
\draw [arrow] (partition) -- (optimize);
\draw [arrow] (optimize) -- (prune);
\draw [arrow] (prune) -- (partition);
\draw [arrow] (prune) -- (end);
\end{tikzpicture}
\Description{A flowchart with parameterize, partition, optimize, and prune}
\caption{COGNAC's overall workflow}
\label{fig:workflow}
\end{figure}

\Cref{fig:workflow} shows a high-level overview of COGNAC's strategy, with the bulk of the work being done in the optimization step.
This step is most effective with relatively small parameterized circuits, hence the need for pre-optimization parameterization and partitioning.

\tikzstyle{phaselabel} = [yshift=1mm]
\tikzstyle{lowerphaselabel} = [yshift=-5mm]

\begin{figure}
\begin{subfigure}{\textwidth}
\centering
\begin{quantikz}[row sep=2mm]
\qw & \gate{H} & \ctrl[wire style={"CZ"}]{1} & \qw & \ctrl[wire style={"CZ"}]{1} & \gate{H} & \qw \\
\qw & \qw & \control{} & \gate{R_x(\pi/4)} & \control{} & \qw & \qw
\end{quantikz}
\caption{Input}
\label{fig:input}
\end{subfigure}
\begin{subfigure}{\textwidth}
\vspace{5mm}
\centering
\begin{quantikz}[row sep=2mm,column sep=1mm]
\qw & \phase[label style=phaselabel]{R_z(\pi/2)} & \gate{R_x(\pi/2)} & \phase[label style=phaselabel]{R_z(\pi/2)} & \ctrl[wire style={"R_{zz}(\pi/2)"}]{1} & \phase[label style=phaselabel]{R_z(-\pi)} & \gate{R_x(0)} & \phase[label style=phaselabel]{R_z(\pi/2)} & \ctrl[wire style={"R_{zz}(\pi/2)"}]{1} & \phase[label style=phaselabel]{R_z(0)} & \gate{R_x(\pi/2)} & \phase[label style=phaselabel]{R_z(\pi/2)} & \qw \\
\qw & \phase[label style=lowerphaselabel]{R_z(-\pi/2)} & \gate{R_x(0)} & \phase[label style=lowerphaselabel]{R_z(\pi/2)} & \control{} & \phase[label style=lowerphaselabel]{R_z(-\pi/2)} & \gate{R_x(\pi/4)} & \phase[label style=lowerphaselabel]{R_z(0)} & \control{} & \phase[label style=lowerphaselabel]{R_z(-\pi)} & \gate{R_x(0)} & \phase[label style=lowerphaselabel]{R_z(\pi/2)} & \qw
\end{quantikz}
\caption{Translated}
\label{fig:translated}
\end{subfigure}
\begin{subfigure}{\textwidth}
\vspace{5mm}
\centering
\begin{quantikz}[row sep=2mm,column sep=2mm]
\qw & \phase[label style=phaselabel]{R_z(\theta_1)} & \gate{R_x(\theta_2)} & \phase[label style=phaselabel]{R_z(\theta_3)} & \ctrl[wire style={"R_{zz}(\theta_{10})"}]{1} & \phase[label style=phaselabel]{R_z(\theta_4)} & \gate{R_x(\theta_5)} & \phase[label style=phaselabel]{R_z(\theta_6)} & \ctrl[wire style={"R_{zz}(\theta_{11})"}]{1} & \phase[label style=phaselabel]{R_z(\theta_7)} & \gate{R_x(\theta_8)} & \phase[label style=phaselabel]{R_z(\theta_9)} & \qw \\
\qw & \phase[label style=lowerphaselabel]{R_z(\theta_{12})} & \gate{R_x(\theta_{13})} & \phase[label style=lowerphaselabel]{R_z(\theta_{14})} & \control{} & \phase[label style=lowerphaselabel]{R_z(\theta_{15})} & \gate{R_x(\theta_{16})} & \phase[label style=lowerphaselabel]{R_z(\theta_{17})} & \control{} & \phase[label style=lowerphaselabel]{R_z(\theta_{18})} & \gate{R_x(\theta_{19})} & \phase[label style=lowerphaselabel]{R_z(\theta_{20})} & \qw
\end{quantikz}
\caption{Parameterized}
\label{fig:parameterized}
\end{subfigure}
\begin{subfigure}{\textwidth}
\vspace{5mm}
\centering
\begin{quantikz}[row sep=2mm,column sep=1mm]
\qw & \phase[label style=phaselabel]{R_z(\texttt{1.57})} & \gate{R_x(\texttt{1.57})} & \phase[label style=phaselabel]{R_z(\texttt{1.57})} & \ctrl[wire style={"R_{zz}(0.79)"}]{1} & \phase[label style=phaselabel]{R_z(\texttt{1.57})} & \gate{R_x(\texttt{0.0})} & \phase[label style=phaselabel]{R_z(\texttt{0.0})} & \ctrl[wire style={"R_{zz}(0.0)"}]{1} & \phase[label style=phaselabel]{R_z(\texttt{0.0})} & \gate{R_x(\texttt{1.57})} & \phase[label style=phaselabel]{R_z(\texttt{1.57})} & \qw \\
\qw & \phase[label style=lowerphaselabel]{R_z(\texttt{1.57})} & \gate{R_x(\texttt{1.57})} & \phase[label style=lowerphaselabel]{R_z(\texttt{1.57})} & \control{} & \phase[label style=lowerphaselabel]{R_z(\texttt{1.57})} & \gate{R_x(\texttt{1.57})} & \phase[label style=lowerphaselabel]{R_z(\texttt{0.0})} & \control{} & \phase[label style=lowerphaselabel]{R_z(\texttt{0.0})} & \gate{R_x(\texttt{0.0})} & \phase[label style=lowerphaselabel]{R_z(\texttt{1.57})} & \qw
\end{quantikz}
\caption{Optimized}
\label{fig:optimized}
\end{subfigure}
\begin{subfigure}{\textwidth}
\vspace{5mm}
\centering
\begin{quantikz}[row sep=2mm,column sep=2mm]
\qw & \phase[label style=phaselabel]{R_z(\texttt{1.57})} & \gate{R_x(\texttt{1.57})} & \phase[label style=phaselabel]{R_z(\texttt{1.57})} & \ctrl[wire style={"R_{zz}(0.79)"}]{1} & \phase[label style=phaselabel]{R_z(\texttt{1.57})} & \gate{R_x(\texttt{1.57})} & \phase[label style=phaselabel]{R_z(\texttt{1.57})} & \qw \\
\qw & \phase[label style=lowerphaselabel]{R_z(\texttt{1.57})} & \gate{R_x(\texttt{1.57})} & \phase[label style=lowerphaselabel]{R_z(\texttt{1.57})} & \control{} & \phase[label style=lowerphaselabel]{R_z(\texttt{1.57})} & \gate{R_x(\texttt{1.57})} & \phase[label style=lowerphaselabel]{R_z(1.)} & \qw
\end{quantikz}
\caption{Pruned}
\label{fig:pruned}
\end{subfigure}
\caption{COGNAC optimization of an inefficient $R_{xx}(\pi/4)$ gate implementation}
\Description{A quantum circuit undergoing parametrization and pruning}
\label{fig:stages}
\end{figure}

As a simplified example, \Cref{fig:stages} shows a circuit that passes through several of these stages.
We discuss each in turn.

\paragraph{Parameterize}
We first translate the input circuit gate-by-gate into a parameterized gate set, here using the gates from \Cref{fig:gateset}.
This is straightforward with existing techniques; \Cref{fig:parameterize} shows how this can be done step by step.
Ignoring the global phase, we first convert all the $CZ$ gates to $R_{zz}$ and $R_z$ gates (\Cref{fig:translate2}).
Then we replace each single-qubit sequence (including blank wires) with a $U_3$ gate (\Cref{fig:translate3}, which is a three-parameter gate that can optimally implement any single-qubit unitary~\cite{qasm}.
The three parameters of the $U_3$ gate correspond to the parameters of an equivalent sequence of $R_z$ and $R_x$ gates through the equivalence $U_3(\theta, \phi, \lambda) \propto R_z(\phi+\frac \pi 2) R_x(\theta) R_z(\lambda - \frac \pi 2)$ (\Cref{fig:translate1}).
We temporarily \emph{increase} the number of single-qubit gates with the goal of creating more opportunities for the later elimination of two-qubit gates.

\begin{figure}
\begin{subfigure}{\textwidth}
\centering
\begin{quantikz}[row sep=2mm]
\qw & \gate{H} & \ctrl[wire style={"CZ"}]{1} & \qw & \ctrl[wire style={"CZ"}]{1} & \gate{H} & \qw \\
\qw & \qw & \control{} & \gate{R_x(\pi/4)} & \control{} & \qw & \qw
\end{quantikz}
\caption{Input}
\end{subfigure}
\begin{subfigure}{\textwidth}
\vspace{5mm}
\centering
\begin{quantikz}[row sep=2mm]
\qw & \gate{H} & \ctrl[wire style={"R_{zz}(\pi/2)"}]{1} & \phase[label style=phaselabel]{R_z(-\pi/2)} & \qw & \ctrl[wire style={"R_{zz}(\pi/2)"}]{1} & \phase[label style=phaselabel]{R_z(-\pi/2)} & \gate{H} & \qw \\
\qw & \qw & \control{} & \phase[label style=lowerphaselabel]{R_z(-\pi/2)} & \gate{R_x(\pi/4)} & \control{} & \phase[label style=lowerphaselabel]{R_z(-\pi/2)} & \qw & \qw
\end{quantikz}
\caption{Two-qubit gates translated}
\label{fig:translate2}
\end{subfigure}
\begin{subfigure}{\textwidth}
\vspace{5mm}
\centering
\begin{quantikz}[row sep=2mm]
\qw & \gate{U_3(\pi/2, 0, \pi)} & \ctrl[wire style={"R_{zz}(\pi/2)"}]{1} &[1cm] \gate{U_3(0, 0, -\pi/2)} & \ctrl[wire style={"R_{zz}(\pi/2)"}]{1} &[1cm] \gate{U_3(\pi/2, 0, \pi/2)} & \qw \\
\qw & \gate{U_3(0,0,0)} & \control{} & \gate{U_3(\pi/4,-\pi/2,0)} & \control{} & \gate{U_3(0, 0, -\pi/2)} & \qw
\end{quantikz}
\caption{Universality of the $U_3$ gate}
\label{fig:translate3}
\end{subfigure}
\begin{subfigure}{\textwidth}
\vspace{5mm}
\centering
\begin{quantikz}[row sep=2mm,column sep=1mm]
\qw & \phase[label style=phaselabel]{R_z(\pi/2)} & \gate{R_x(\pi/2)} & \phase[label style=phaselabel]{R_z(\pi/2)} & \ctrl[wire style={"R_{zz}(\pi/2)"}]{1} & \phase[label style=phaselabel]{R_z(-\pi)} & \gate{R_x(0)} & \phase[label style=phaselabel]{R_z(\pi/2)} & \ctrl[wire style={"R_{zz}(\pi/2)"}]{1} & \phase[label style=phaselabel]{R_z(0)} & \gate{R_x(\pi/2)} & \phase[label style=phaselabel]{R_z(\pi/2)} & \qw \\
\qw & \phase[label style=lowerphaselabel]{R_z(-\pi/2)} & \gate{R_x(0)} & \phase[label style=lowerphaselabel]{R_z(\pi/2)} & \control{} & \phase[label style=lowerphaselabel]{R_z(-\pi/2)} & \gate{R_x(\pi/4)} & \phase[label style=lowerphaselabel]{R_z(0)} & \control{} & \phase[label style=lowerphaselabel]{R_z(-\pi)} & \gate{R_x(0)} & \phase[label style=lowerphaselabel]{R_z(\pi/2)} & \qw
\end{quantikz}
\caption{Conversion to single-parameter gates}
\label{fig:translate1}
\end{subfigure}
\caption{A closer look at the initial gate translation step}
\label{fig:parameterize}
\end{figure}

\paragraph{Partition}
The example circuit in \Cref{fig:stages} is small and partitioning is unnecessary. 
However, very large circuits must be divided into smaller subcircuits for the optimization step to be tractable. \Cref{fig:partition} illustrates the desired outcome of this stage with an example five-qubit circuit.
Although we may appear to be reordering gates, we do so only with \emph{independent} gates, so the directed acyclic graph (\textsc{dag}) corresponding to the circuit remains unchanged.
Previous work has explored the use of various partitioning algorithms for quantum circuits \cite{qgo, gtqcp}, and our COGNAC implementation uses the \texttt{QuickPartitioner} from BQSKit \cite{bqskit}.

\begin{figure}
\begin{subfigure}{\linewidth}
\centering
\begin{quantikz}[row sep=3mm,column sep=2mm]
\qw & \gate{} & \ctrl{1} & \gate{} & \ctrl{1} & \gate{} & \qw & \ctrl{4} & \gate{} & \ctrl{4} & \gate{} & \ctrl{3} & \gate{} & \qw & \qw & \qw & \qw \\
\qw & \gate{} & \control{} & \gate{} & \control{} & \gate{} & \ctrl{1} & \qw & \gate{} & \qw & \qw & \qw & \qw & \qw & \ctrl{2} & \gate{} & \qw \\
\qw & \gate{} & \qw & \qw & \qw & \qw & \control{} & \qw & \gate{} & \qw & \qw & \qw & \qw & \ctrl{2} & \qw & \gate{} & \qw \\
\qw & \gate{} & \qw & \qw & \qw & \qw & \qw & \qw & \qw & \qw & \qw & \control{} & \gate{} & \qw & \control{} & \gate{} & \qw \\
\qw & \gate{} & \qw & \qw & \qw & \qw & \qw & \control{} & \gate{} & \control{} & \gate{} & \qw & \qw & \control{} & \gate{} & \qw & \qw
\end{quantikz}
\caption{An example five-qubit circuit}
\end{subfigure}
\begin{subfigure}{\linewidth}
\centering
\begin{quantikz}[row sep=3mm,column sep=2mm]
\qw & \gate{}\gategroup[5,steps=9,style={dashed,rounded corners}]{} & \ctrl{1} & \gate{} & \ctrl{1} & \gate{} & \ctrl{4} & \gate{} & \ctrl{4} & \gate{} &[5mm] \qw & \qw & \qw & \qw & \qw &[5mm] \qw\gategroup[4,steps=5,style={dashed, rounded corners}]{} & \ctrl{3} & \gate{} & \qw & \qw & \qw \\
\qw & \gate{} & \control{} & \gate{} & \control{} & \gate{} & \qw & \qw & \qw & \qw & \qw\gategroup[4,steps=5,style={dashed,rounded corners}]{} & \ctrl{1} & \gate{} & \qw & \qw & \qw & \qw & \qw & \ctrl{2} & \gate{} & \qw \\
\qw & \qw & \qw & \qw & \qw & \qw & \qw & \qw & \qw & \qw & \gate{} & \control{} & \gate{} & \ctrl{2} & \gate{} & \qw & \qw & \qw & \qw & \qw & \qw \\
\qw & \qw & \qw & \qw & \qw & \qw & \qw & \qw & \qw & \qw & \qw & \qw & \qw & \qw & \qw & \gate{} & \control{} & \gate{} & \control{} & \gate{} & \qw \\
\qw & \gate{} & \qw & \qw & \qw & \qw & \control{} & \gate{} & \control{} & \gate{} & \qw & \qw & \qw & \control{} & \gate{} & \qw & \qw & \qw & \qw & \qw & \qw
\end{quantikz}
\caption{Three 3-qubit windows that can each be optimized independently}
\end{subfigure}
\caption{Partitioning a five-qubit circuit into three-qubit windows}
\label{fig:partition}
\end{figure}

\paragraph{Optimize}
Each window consists essentially of two components: the gate structure (or \textit{ansatz}) and the real parameter vector $\vec{\theta} = \langle \theta_1, \ldots, \theta_n \rangle \in \mathbb{R}^n$.
The circuit in Figure~\ref{fig:parameterized}, for example, would have $n=20$.

The ansatz defines a matrix-valued function $U(\vec{\theta})$ describing the ideal unitary implemented by the circuit with these parameter values.
If the input parameters are $\vec{\theta}_0$, then there will generally be other possible settings $\vec{\theta}_\star$ such that $U(\vec{\theta}_\star) \approx U(\vec{\theta}_0)$.
Such alternative parameters still approximate the target unitary and may be preferable to $\vec{\theta}_0$.
For example, if $\theta_{11}=0$, then the $R_{zz}(\pi/2)$ gate becomes an $R_{zz}(0)$ gate, which is an identity operator and can be safely removed.

To express these preferences, we can devise a \textit{figure of merit} to be used with gradient ascent.
(This is the opposite of the \textit{cost function} used in gradient \emph{descent}, an equivalent way to frame the problem.)
The standard figure of merit for pulse engineering~\cite{quantum-control} would be $|\trace(U(\vec{\theta}_0)^\dagger U'(\vec{\theta}_\star))|$, assuming that $U'(\cdot)$ accounts for noise.

For simplicity, we model the noise with
\[
U'(\vec{\theta}_\star) = \Fid(\vec{\theta}_\star)U(\vec{\theta}_\star).
\]
Here, $\Fid : \mathbb{R}^n \to [0, 1]$ is a differentiable fidelity function that we will define more precisely in Section~\ref{sec:merit}.
For now, just note that it decays approximately exponentially with the sum of all the two-qubit rotation angles $\theta$.
This model -- effectively depolarizing noise proportional to the size of the circuit\footnote{A depolarizing noise model assumes that the quantum state decays into a maximally mixed state with probability $1 - \Fid(\vec{\theta}_\star)$, which increases with the duration of the circuit's entangling gates.} -- is simpler and less precise than the noise models typically used in pulse engineering, but it has the advantage that it is fast to compute and differentiate while still general enough to be useful on a variety of quantum hardware.

An existing gradient-based optimization algorithm (in our case L-BFGS~\cite{lbfgs}) can then adjust the parameters to maximize the figure of merit, using the gradient to inform its search direction.
After many iterations, it reaches a local maximum value for the figure of merit and updates the circuit with the appropriate parameter values (\Cref{fig:optimized}).

\paragraph{Prune}
The pruning stage is fairly straightforward, removing gates whose corresponding unitary is (approximately) the identity operator and producing the smaller circuit in \Cref{fig:pruned}.
After pruning, we can repartition the circuit and repeat the process.
We found the best performance when running multiple optimization rounds with increasing window sizes, for example, 3, then 4, then 5.
Because the pruning stage alters the circuit \textsc{dag}, it can enable more expansive windows in successive optimization rounds.

The two-qubit example in \Cref{fig:stages} is small enough that existing optimizers could optimally synthesize the entire two-qubit circuit, but COGNAC has the advantage of being much more general: It readily applies to larger circuits without changing the gate set or connectivity constraints.

\section{Implementation}
\label{sec:impl}

We have implemented COGNAC as an open source Qiskit compiler plugin.
It uses BQSKit's partitioning scheme to divide a quantum circuit into smaller subcircuits, and then relies on TensorFlow's implementation of the L-BFGS quasi-Newton optimization algorithm~\cite{tensorflow,lbfgs}.
Most of the interesting implementation work is thus in constructing a differentiable cost function (or figure of merit) that a GPU can efficiently evaluate.

\subsection{Figure of Merit}
\label{sec:merit}
COGNAC is designed to run relatively late in the compilation process, after gates have been decomposed into a native gate set and logical qubits have been mapped to the hardware layout.
COGNAC then optimizes the subcircuits in which all two-qubit gates are parameterized.

\begin{figure}
\fboxsep0pt
\begin{subfigure}{0.37\linewidth-1pc}
\begin{align*}
U'(\vec{\theta}) &= \Fid(\vec{\theta})U(\vec{\theta}) \\
\Fid(\vec{\theta}) &\defeq \Pi_{\theta_i \in \vec{\theta}} (1 - E_i \cdot \xi(\ell(\theta_i))) \\
\xi(x) &\defeq \frac{x(3-x)}{2}
\end{align*}
\end{subfigure}
\hfill
\begin{subfigure}{0.63\linewidth-1pc}
\includegraphics{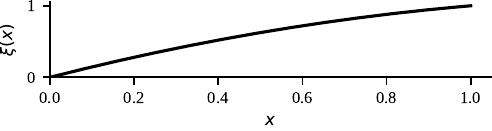}
\end{subfigure}
\caption{Our simplified error model}
\label{fig:xi}
\end{figure}

As mentioned earlier, we use a standard figure of merit, $|\trace(U(\vec{\theta}_0)^\dagger U'(\vec{\theta}_\star))|$, assuming that $U$ is an idealized unitary for the target operator and $U'$ approximates noise.
Each of these is calculated as the product of a series of matrices, each corresponding to a gate in the circuit.
With $U$, these matrices are those of Figure~\ref{fig:gateset}, but for $U'$, we account for noisy gates using the model in \Cref{fig:xi}.
It defines circuit fidelity $\Fid$ as the product of the fidelity of each individual gate in the circuit.
Here, $E_i$ is a hardware-dependent constant that describes the estimated error rate of the $i$\textsuperscript{th} gate in the circuit, for example 0.01 when $\theta_i$ is the parameter for a gate with 99\% fidelity.
The $\ell$ function is a periodic piecewise-linear function that converts the rotation angle to the duration of the underlying hardware pulse.
The function $\xi$, depicted in Figure~\ref{fig:xi}, describes how the error scales with the duration of the gate.
This definition ensures that $\xi(1)=1$ and $\xi(0) = 0$ (accounting for the presence or absence of a fully entangling gate), and its positive derivative encourages the optimizer to follow the gradient to a zero-duration gate.
Leaving those specifications aside, our $\xi$ is somewhat arbitrary.
Picking $\xi(x)=x$ or $\xi(x)=\sin^2\left(\frac{\pi}{2}x\right)$ would have been just as valid and may model certain hardware with greater accuracy.
In practice, we find that these functions lead to circuits with a large number of small-angle gates, and by making the gradient steeper around $\theta \approx 0$, we encourage more of these angles to drop to $\theta = 0$.

The function $U(\cdot)$ is a linear function of the sines and cosines of the various rotation angles, and the noise model adds a quadratic multiplier, so computing the gradient of our figure of merit is straightforward, and TensorFlow can compute this automatically.
We implement all of this calculation using TensorFlow, which evaluates the gradient many times as it searches for better parameters.

One side effect of tracking the error per gate is that it becomes possible to use different values of $E$ for the different hardware gates.
On systems that report the fidelities of their individual two-qubit gates, this allows us to encourage the optimizer to prioritize shortening those gates which are especially noisy, in effect doing more computation on the higher-fidelity qubits.

\subsection{GPU Acceleration}
COGNAC's approach is particularly well-suited to GPU acceleration.
Evaluating the cost function is largely a matter of matrix multiplication, a task at which GPUs excel.
Once we define a cost function, TensorFlow's automatic differentiation functionality can compute the gradient as well.

We use TensorFlow's implementation of the L-BFGS algorithm for the quasi-Newton numerical optimization.
Because COGNAC's optimization stage is entirely implemented (and itself optimized) with Tensorflow, it is essentially one large computational job that can be offloaded to the GPU, minimizing the (relatively slow) data transfers between CPU and GPU.
Both COGNAC and BQSKit rely on L-BFGS for their optimization, but while BQSKit iteratively removes gates and tries to find satisfactory parameters for the remaining gates, COGNAC can optimize an entire window at once.
BQSKit was not designed for GPU acceleration, and it is questionable whether their strategy is compatible with this kind of hardware accelerator, given its sequential core.
COGNAC is also parallelized across the various windows, though GPU memory limitations require some batching of the largest circuits.

\section{Evaluation}
\label{sec:evaluation}
To evaluate COGNAC, we use it to optimize a range of benchmark circuits, which we then run on quantum hardware, comparing the Hellinger fidelity to that achieved by other optimizers.

\paragraph{Research questions}
We try to answer the following research questions:
\begin{itemize}
\item (\ref{sec:comparison}) How does COGNAC compare with existing compiler optimizations, both in terms of classical compilation time and experimental fidelity on quantum hardware?
\item (\ref{sec:window}) How is COGNAC's performance affected by the size of the optimization windows?
\end{itemize}

\paragraph{Optimizers}
We compare COGNAC's performance on benchmark circuits with that of three existing quantum circuit optimizers: Qiskit~\cite{qiskit}, TKET~\cite{tket}, and BQSKit~\cite{bqskit}.

In general, we use these optimizers with default settings.
All include an \optlev{} parameter analogous to the \texttt{-O} flags in GCC, and Qiskit is the only one that we modify, being faster than any other optimizer even with the highest setting (\texttt{\optlev=3}).
At \texttt{\optlev=3}, TKET does not support some of the instructions in our benchmark circuits,\footnote{In particular, it ignores optimization \textit{barriers} between quantum gates and measurements, allowing it to find optimizations that significantly alter the underlying unitary, which are outside the scope of this comparison.} so we use it with the default level of \texttt{2}.

We call the Qiskit \texttt{transpile} function with arguments that disable the routing stage, as our pre-processed benchmark circuits are already properly routed to the target architecture.
The standard TKET compilation function does not provide such a compiler flag, and this poses a problem for our comparison, since the inclusion of this routing stage usually \emph{worsens} a circuit which is already routed.\footnote{This is for two main reasons. The first is that TKET's routing stage does not properly process the noise data from IBM's machines, leading it to select lower-fidelity qubits. The second is that TKET performs some prerouting optimization under the assumption that routing does not matter, but these optimizations are outweighed by the subsequent increase in gate count when the routing stage must then insert corrective \textsc{swap} gates.}
For this reason, we write our own compilation pass sequence that strips out the routing stage and prevents any optimization from introducing virtual \textsc{swap} gates.
We include data from the default (routing-inclusive) TKET compilation in \Cref{app:tables}.
BQSKit similarly does not provide an easy way to disable routing, but it does not reroute as aggressively as TKET and does not end up modifying qubit placement for our benchmark circuits.

We run BQSKit with the default window size of 3 qubits and COGNAC with a window size of 5 qubits.
For both of these optimizers, a larger window size means more optimization but a longer compile time.
However, as we will see in \Cref{sec:comparison}, BQSKit typically requires more time with window size 3 than COGNAC requires with window size 5, so we feel that this is a fair comparison.

\paragraph{Benchmarks}
We evaluate each optimizer's performance across a variety of benchmark circuits from the MQT~Bench collection~\cite{mqt}, a diverse benchmark suite that includes a range of algorithms and circuit components, such as Grover's algorithm, a quantum walk, and a variational quantum eigensolver.
We preprocess each circuit to adhere to the architecture constraints of our target hardware (IBM~Torino), producing an initial circuit that can run on the target machine, which we can directly compare with the optimized circuits.
It also allows for a more controlled comparison, ignoring the ways in which different compilers \textit{route} virtual qubits to physical qubits and limiting the evaluation to post-routing optimizations.
We preprocess circuits using Qiskit with \texttt{\optlev=3} for the routing stage and \texttt{\optlev=1} for all other stages of compilation.
This ensures that the circuit is mapped to high-fidelity physical qubits but still has plenty of opportunities for optimization. 

MQT~benchmarks come in multiple sizes, only a few of which are useful for an optimization comparison.
For example, there is very little room for improving the two-qubit circuits, which are already close to optimal, with fidelity greater than 99\% in general.
At the other end of the spectrum, very large circuits with very low fidelity are similarly uninteresting.
For each benchmark, we classically simulated preprocessed circuits in a variety of sizes on a noisy model of the quantum hardware, selecting the largest size that achieved a simulated fidelity of at least $\frac{1}{e} \approx 37\%$, based on IonQ's threshold for their test of \textit{algorithmic qubits}~\cite{aq}.
Of the 28 benchmarks included in MQT~Bench, Shor's algorithm is the only one for which even the smallest size (over 60,000 gates) is too large to run with reasonable fidelity, so we omit it from our benchmarks.

\paragraph{Hardware specifications}
We ran the optimizers on a four-core 3.2 GHz machine with 32 GB of memory.
This (classical) computer was equipped with an NVIDIA GeForce RTX 2080 Ti GPU.
We submitted our optimized circuits to one of IBM's ``Heron'' quantum computers: \texttt{ibm\_torino}, a device with 133 superconducting qubits.
Error rates vary by qubit, but at the time of the tests, the reported median CZ error rate was approximately $4\times 10^{-3}$.

\subsection{Performance comparison}
\label{sec:comparison}

\begin{table}
\caption{Experimental Hellinger fidelity from various optimizers}
\label{tab:fidelity}
\begin{tabular}{@{}rlrrrrr@{}}\toprule
&&\multicolumn{5}{c}{Hellinger fidelity} \\ \cmidrule(l){3-7}\\
Benchmark & \# qubits & Unoptimized & Qiskit & TKET & BQSKit & COGNAC \\ \midrule
ae & 11 & 20\% & 19\% & 18\% & 21\% & \textbf{32\%} \\
dj & 22 & 17\% & 15\% & \textbf{19\%} & 14\% & 17\% \\
ghz & 30 & 31\% & \textbf{34\%} & 31\% & 29\% & 33\% \\
graphstate & 8 & \textbf{98\%} & \textbf{99\%} & \textbf{98\%} & \textbf{98\%} & \textbf{98\%} \\
groundstate & 4 & 88\% & 88\% & 87\% & 87\% & \textbf{89\%} \\
grover-noancilla & 5 & 29\% & 30\% & 29\% & 27\% & \textbf{59\%} \\
grover-v-chain & 5 & 59\% & 59\% & 59\% & 61\% & \textbf{78\%} \\
portfolioqaoa & 8 & \textbf{67\%} & 67\% & 66\% & 66\% & \textbf{67\%} \\
portfoliovqe & 8 & 57\% & 58\% & \textbf{62\%} & 57\% & 59\% \\
pricingcall & 7 & 36\% & 38\% & 28\% & 46\% & \textbf{51\%} \\
pricingput & 7 & 36\% & 34\% & 23\% & 44\% & \textbf{56\%} \\
qaoa & 12 & 84\% & \textbf{85\%} & 79\% & 84\% & 84\% \\
qft & 8 & 98\% & 98\% & 95\% & 97\% & \textbf{98\%} \\
qftentangled & 8 & 88\% & 87\% & 87\% & 88\% & \textbf{88\%} \\
qnn & 9 & 80\% & \textbf{81\%} & 75\% & 76\% & 77\% \\
qpeexact & 10 & 29\% & 24\% & 20\% & 34\% & \textbf{35\%} \\
qpeinexact & 12 & 23\% & 22\% & 13\% & 13\% & \textbf{37\%} \\
qwalk-noancilla & 4 & 40\% & 39\% & 40\% & 43\% & \textbf{78\%} \\
qwalk-v-chain & 5 & 40\% & 43\% & 37\% & 44\% & \textbf{67\%} \\
random & 8 & 86\% & 86\% & 84\% & 85\% & \textbf{88\%} \\
realamprandom & 8 & 64\% & 63\% & 65\% & 65\% & \textbf{68\%} \\
routing & 12 & \textbf{70\%} & 64\% & 70\% & 67\% & 61\% \\
su2random & 8 & 79\% & 78\% & \textbf{80\%} & 78\% & \textbf{80\%} \\
tsp & 9 & \textbf{91\%} & 91\% & 85\% & \textbf{92\%} & \textbf{92\%} \\
twolocalrandom & 9 & 58\% & 58\% & 56\% & 59\% & \textbf{62\%} \\
vqe & 16 & \textbf{58\%} & 57\% & 57\% & \textbf{58\%} & 57\% \\
wstate & 30 & \textbf{21\%} & 20\% & 19\% & \textbf{21\%} & \textbf{21\%} \\
\bottomrule
\end{tabular}
\end{table}

\paragraph{How does COGNAC compare with existing compiler optimizations, both in terms of classical compilation time and experimental fidelity on quantum hardware?}
In our experiments on IBM's superconducting hardware, we found that COGNAC often outperforms existing optimizers.
We optimized the benchmark circuits using the optimizers and measured how well the output probability distribution (based on 10,000 samples) matched the ideal target distribution.
We measure experimental performance using \textit{Hellinger fidelity}, a measure of similarity between probability distributions commonly used to evaluate noisy quantum programs~\cite{superstaq,aq}.
For each circuit, we first run a noiseless classical simulation to determine the ideal output distribution, and then we compare this with the actual measurement results.
The Hellinger fidelity between two probability mass functions $p$ and $q$ is a value ranging from 0 to 1, calculated as:
\[
\left(\sum_x \sqrt{p(x)q(x)}\right)^2.
\]

\Cref{tab:fidelity} shows the results with probabilities rounded to the nearest percentage point.
This table uses COGNAC with five-qubit windows; we discuss the effect of window size in \Cref{sec:window}.
COGNAC outperforms all other optimizers on 17 of the 27 benchmarks, sometimes by a significant margin.

\begin{figure}
\includegraphics{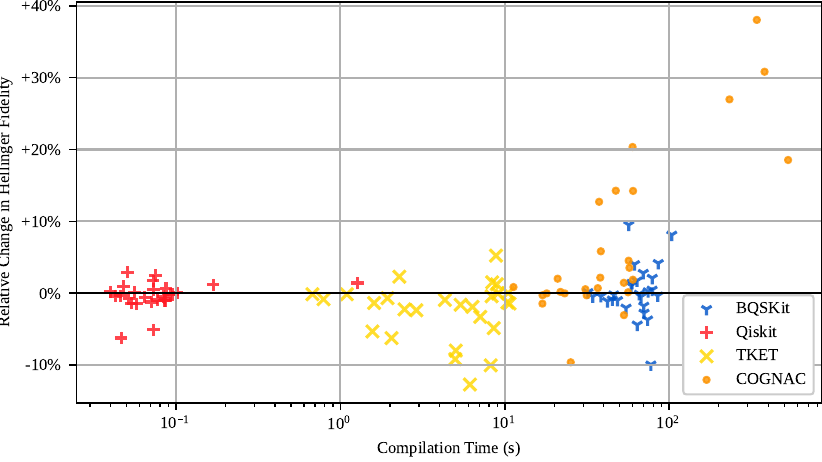}
\Description{A scatterplot comparing compilation time with relative change in Hellinger fidelity across various optimizers. Qiskit has the shortest time, followed by TKET. The points for BQSKit and COGNAC are more intermingled, but with COGNAC's points generally higher in fidelity and shorter in time.}
\caption{Fidelity vs. Time for Various Optimizers}
\label{fig:comparison}
\end{figure}

It achieves these numbers while remaining competitive in terms of compilation time.
\Cref{fig:comparison} shows the trade-off between compile-time performance versus run-time performance, with optimization time plotted on the $x$ axis and relative change in fidelity (subtracting the unoptimized fidelity) on the $y$ axis.
Qiskit and TKET are significantly faster than COGNAC in general, but BQSKit usually requires more time to compile than COGNAC.
In total, COGNAC was faster to compile than BQSKit on 20 of the 27 benchmarks.
COGNAC outperformed BQSKit in terms of \emph{both} experimental fidelity \emph{and} compile time on 16 of the benchmarks.

\subsection{Window size}
\label{sec:window}

\paragraph{How is COGNAC's performance affected by the size of the optimization windows?}
COGNAC's window size is the main parameter to adjust its behavior.
Larger windows create more opportunities for optimizing circuits, but this comes at the cost of additional computation.
Each additional qubit quadruples the memory requirements of computing the figure of merit, and that is before accounting for the impact of additional gate parameters.

\begin{figure}
\includegraphics{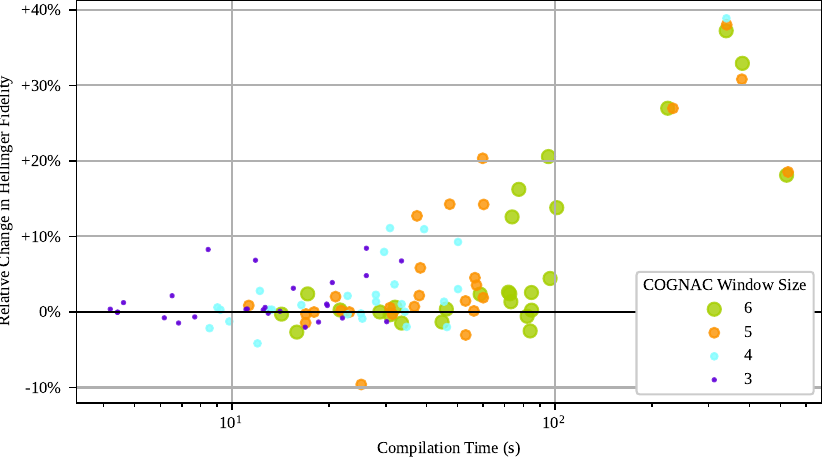}
\caption{Fidelity vs. Time for Various COGNAC Window Sizes}
\label{fig:window}
\end{figure}

\Cref{fig:window} visualizes the impact of window size on Hellinger fidelity and compilation time.
The optimization here is cumulative; window size 5 actually means three rounds of optimization: one with window size 3, another with window size 4, and a final one with window size 5.
(This is true for all of our experimental results.)
We found that this strategy achieves the best performance, as optimizing with the smaller windows is faster and can prune the circuit \textsc{dag} in a way that allows the larger windows to include more gates.

As seen in \Cref{fig:window}, larger and larger window sizes begin to yield diminishing returns.
On most of these benchmarks, window sizes 5 and 6 are essentially tied for fidelity, as the 6-qubit optimization windows find few opportunities for optimization in the (already heavily optimized) circuits.

\section{Discussion}

So far, we have not addressed the obvious question: Can we use COGNAC \emph{in conjunction} with other optimizers?
After all, the most widely used classical compilers are complex toolchains with a lot of different kinds of optimization, and perhaps the different optimizers in our comparison have different strengths that can complement each other.

\begin{figure}
\includegraphics{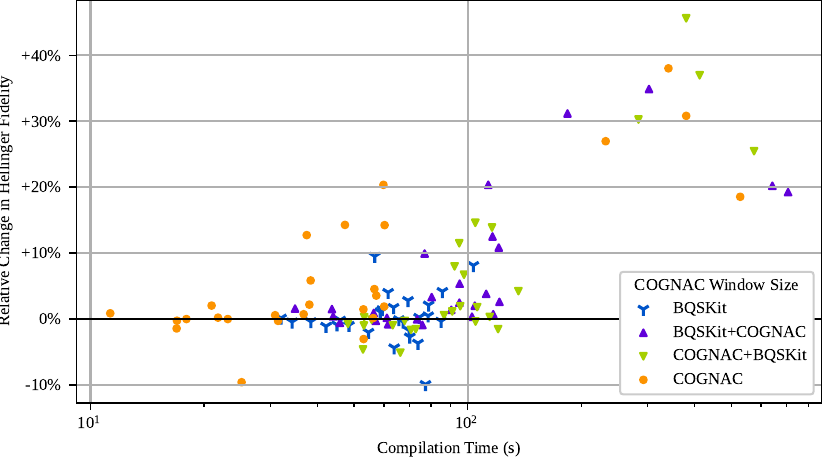}
\caption{Fidelity vs. Time for BQSKit \emph{with} COGNAC}
\label{fig:combo}
\end{figure}

We do not yet have a definitive answer to this question.
\Cref{fig:combo} shows the results of combining COGNAC with BQSKit, with COGNAC first (``COGNAC+BQSKit'') or BQSKit first (``BQSKit+COGNAC'').
Although the combination of the two sometimes achieves greater fidelity than either of the two individually, it comes (predictably) at the cost of a longer runtime.
More interesting is the fact that COGNAC sometimes achieves its highest fidelity in isolation, with the addition of BQSKit serving more as a hindrance.
More work in this direction is needed if we hope to use COGNAC to its full potential. 

Our experimental evaluation has also been limited to IBM's superconducting hardware, but COGNAC can be adapted to work with any parameterized gate set.
For example, IonQ's trapped ion quantum computers provide native $R_{xx}$ gates, and COGNAC's versatile optimization strategy finds improvements to these circuits as well.

\section{Related Work}
COGNAC attempts to blur the line between compiler optimization and pulse engineering, and developments from both of these research areas have influenced its principles.

Gate-level optimizers, like optimizers for classical programs, tend to rely on rewriting rules.
A notable exception is in the synthesis of two-qubit gates, for which there is a general procedure that is known to be optimal~\cite{two-qubit-decomposition}.
TKET applies this procedure to every two-qubit gate in a circuit, reducing the number of consecutive gates applied to the same two qubits~\cite{tket}.
This synthesis has been improved by accounting for noise and including fractional-angle gates but is still limited to two-qubit gates.

Some researchers have noted that there are a number of different gate sets for different quantum hardware and that rewrite rules necessarily make assumptions about gate sets that may not apply.
One solution to this problem has been the automatic generation of large numbers of rewrite rules for an arbitrary gate set, as done by Quartz~\cite{quartz} and Queso~\cite{queso}.
Quartz is notable for its guarantee of generating \emph{all} possible rewrite rules of a given size, but requires considerable computational resources.
Queso represents gate parameters as symbolic angles, allowing them to generate rewrite rules like ``$R_z(\alpha)R_z(\beta) \mapsto R_z(\alpha + \beta)$.''
However, the symbolic transformations are limited to relatively simple formulas, and neither of these systems considers \emph{approximate} rewrites.

Quarl~\cite{quarl} is another recent quantum circuit optimizer that builds on Quartz's generated rewrite rules, using reinforcement learning on a neural network to decide which rewrites to perform.
Unfortunately, it is designed to run for hours on a supercomputer, so it is impractical as part of a typical compiler toolchain.
Like Queso and Quartz, Quarl is limited to idealized equivalences and does not allow for approximations.

In the quantum control community, GRAPE (gradient ascent pulse engineering) has emerged as a popular tool for designing high-fidelity pulse sequences~\cite{qoc}.
Like COGNAC, GRAPE uses numerical optimization, iteratively adjusting parameters to maximize fidelity to a target operation.
However, operating at the level of hardware pulses, GRAPE has a very large parameter space even for small operations and is not scalable~\cite{cqe}.
GRAPE can effectively optimize the hardware pulses that implement the one- and two-qubit basis gates of a machine, but its pulses are only as accurate as its physical model of the quantum system, and noise can be difficult to precisely characterize on practical hardware~\cite{grape-protocol}.
Practical applications of GRAPE to produce well-calibrated gates tend to involve an iterative protocol involving feedback from hardware experiments~\cite{feedback-grape,d-grape}, which makes it unsuitable as part of a general-purpose compiler optimization.
COGNAC aims fittingly to distill the key features of GRAPE while porting it to a different context.

Among existing tools, BQSKit~\cite{bqskit,qgo,quest,leap} is the most comparable to COGNAC.
Like COGNAC, it uses L-BFGS numerical optimization.
However, their fundamentally sequential algorithm iterates through each gate in a window.
For each gate, they try to remove it and then use numerical optimization to adjust the parameters of the remaining gates.
This optimization uses an idealized figure of merit (similar to setting $\Fid = 1$ in our model), and gate removal is successful if the result is within a specified tolerance (\texttt{synthesis\_epsilon}, by default $10^{-8}$).
This means that while it is parallelizable across windows, each window's algorithm must act sequentially and cannot run on a GPU, limiting it to smaller window sizes.
Its design is also not as adaptive to different levels of machine noise; the \texttt{synthesis\_epsilon} parameter is a global constant and it is not clear how it might be adjusted with machine noise.
Although effectively employing some of the tools of pulse engineering, BQSKit does not take full advantage of the possibilities for noise modeling and larger-scale optimization.

\section{Conclusion and Future Work}

In this work, we presented COGNAC, a novel tool that draws on techniques from pulse engineering to apply them directly to quantum optimizers based on continuous gate sets. We saw that COGNAC usually outperforms the state-of-the-art BQSKIT compiler, though which tool developers choose to use in practice will differ based on a variety of factors. These factors include the size of the circuit, which relates to the compilation time, as well as the desired fidelity. We saw that COGNAC itself can improve the fidelity of its output through larger window sizes, though this has diminishing returns and comes at the cost of increased compilation time.

COGNAC has some limitations that future work could try to handle.
An obvious one is the fixed input ansatz; COGNAC does not allow gates to be reordered or applied to different qubits.
Future optimizers could involve additional compilation passes designed to complement COGNAC by reordering gates or adding new ones.
Quarl-style reinforcement learning could be useful here and for improving other elements of COGNAC, like its cost function and its process for selecting optimization windows in larger circuits.

In future work, we hope to apply these optimizations to additional domains and combine them with existing tools. Following Littiken et al.~\cite{litteken2023dancing} and Liu et al.~\cite{liu2025direct}, hardware vendors could provide native parameterized three-qubit gates, to which we could apply COGNAC-style optimizations, potentially to great benefit. We could also apply COGNAC to novel devices, such as those with higher-dimensional states (qudits) or continuous variable quantum computing, as used on photonic quantum computers. COGNAC's techniques may even prove to have applications within classical computing, as it demonstrates the costs of discretization when compute is at a premium.

\section*{Data-Availability Statement}
All of our work will be made publicly available. We intend to submit an
artifact for artifact evaluation that includes the implementation of
COGNAC, as well as scripts to re-execute the experiments carried out 
when possible. Note that many of the experiments were carried out on 
IBM's quantum hardware.

\begin{acks}
This research used resources of the Oak Ridge Leadership Computing Facility, which is a DOE Office of Science User Facility supported under Contract DE-AC05-00OR22725.
This material is supported by the Air Force Office of Scientific Research under award number FA95502310406.
We thank Murphy Yuezhen Niu, Yuxiang Peng, Michael Hicks, and Benjamin Quiring for their useful feedback on an earlier draft of this report.
\end{acks}

\bibliographystyle{plainurl}
\bibliography{references}

\appendix
\section{Additional tables}
\label{app:tables}

\Cref{tab:unmodified-tket} illustrates our rationale for manually removing TKET's routing stage, as explained in \Cref{sec:evaluation}.
This table contains the data that would be in \Cref{tab:fidelity} if we used TKET's default compilation method.
As can be seen in the many poor low values, any positive effect of TKET's optimizations is overwhelmed by the negative effects of poor qubit mapping.
\begin{table}
\caption{Experimental Hellinger fidelity with unmodified TKET}
\label{tab:unmodified-tket}
\begin{tabular}{@{}rlrr@{}}\toprule
&&\multicolumn{2}{c}{Hellinger fidelity} \\ \cmidrule(l){3-4}\\
Benchmark & \# qubits & Unoptimized & TKET \\ \midrule
ae & 11 & 20\% & 10\% \\
dj & 22 & 17\% & 4\% \\
ghz & 30 & 31\% & 32\% \\
graphstate & 8 & 98\% & 98\% \\
groundstate & 4 & 88\% & 84\% \\
grover-noancilla & 5 & 29\% & 21\% \\
grover-v-chain & 5 & 59\% & 58\% \\
portfolioqaoa & 8 & 67\% & 64\% \\
portfoliovqe & 8 & 57\% & 62\% \\
pricingcall & 7 & 36\% & 33\% \\
pricingput & 7 & 36\% & 28\% \\
qaoa & 12 & 84\% & 80\% \\
qft & 8 & 98\% & 97\% \\
qftentangled & 8 & 88\% & 76\% \\
qnn & 9 & 80\% & 74\% \\
qpeexact & 10 & 29\% & 3\% \\
qpeinexact & 12 & 23\% & 12\% \\
qwalk-noancilla & 4 & 40\% & 45\% \\
qwalk-v-chain & 5 & 40\% & 28\% \\
random & 8 & 86\% & 81\% \\
realamprandom & 8 & 64\% & 67\% \\
routing & 12 & 70\% & 57\% \\
su2random & 8 & 79\% & 77\% \\
tsp & 9 & 91\% & 71\% \\
twolocalrandom & 9 & 58\% & 42\% \\
vqe & 16 & 58\% & 59\% \\
wstate & 30 & 21\% & 20\% \\
\bottomrule
\end{tabular}
\end{table}

\end{document}